\documentclass[letterpaper,amsmath,
  amssymb,aps,prd,nofootinbib,
singlecolumn,
superscriptaddress,altaffilsymbol,12pt]{revtex4-1}

\usepackage{amsmath}
\usepackage{amsfonts}
\usepackage{amssymb}
\usepackage{graphicx}
\usepackage{psfrag}
\usepackage[hang]{footmisc}
\usepackage{setspace}
\usepackage{caption}
\usepackage[dvipsnames]{xcolor}
\newcommand{\beq}{\begin{equation}}
\newcommand{\eeq}{\end{equation}}
\newcommand{\bea}{\begin{eqnarray}}
\newcommand{\eea}{\end{eqnarray}}
\newcommand{\beqn}{\begin{equation*}}
\newcommand{\eeqn}{\end{equation*}}
\newcommand{\bean}{\begin{eqnarray*}}
\newcommand{\eean}{\end{eqnarray*}}

\newcommand*{\cref}[1]{Chapter~\ref{#1}}

\usepackage{graphicx}
\usepackage{dcolumn}
\usepackage{psfrag} 
\usepackage{booktabs}

\begin{document}

\title{Measuring the expansion of the universe}

 \author{Gabriel Germ\'an}\email{e-mail: gabriel@icf.unam.mx}
 \affiliation{Instituto de Ciencias F\'{\i}sicas, Universidad Nacional Aut\'onoma de M\'exico, Cuernavaca, Morelos, 62210, Mexico}


\begin{abstract}
We draw a figure from where it is possible to measure the number of e-folds of expansion of the universe with a ruler. We find model independent bounds for the number of e-folds during inflation, reheating and radiation. We also give a lower bound to the size of the  universe at the beginning of observable inflation. Finally, we show that consistency with a relevant diagram requires the existence of a new form of energy to drive the present expansion of the universe.

\end{abstract}

                              
\maketitle


\section {\bf Introduction}\label{Intro}
The idea of the inflationary universe was proposed some 40 years ago \cite{Guth:1980zm}, (for reviews see e.g., \cite{Linde:1984ir}, \cite{Lyth:1998xn}, \cite{Martin:2018ycu}). Throughout these years, a large number of articles has been published investigating different aspects of inflation as well as of the following stage of evolution of the universe, called reheating, where several ways of implementing constraints have been proposed. \cite{Martin:2006rs}-\cite{Ji:2019gfy}, (for reviews on reheating see e.g.,   \cite{Bassett:2005xm},  \cite{Allahverdi:2010xz}, \cite{Amin:2014eta}). One of the important parameters in the study of the evolution of the universe is the number of  e-folds $N_{ij}\equiv \ln\frac{a_j}{a_i}$ which measures its expansion from the time the scale factor was $a_i$ to the time when the scalar factor was $a_j$\footnote{In the Appendix we collect the definitions for the various $N_{ij}$ used throughout the article.} . Sometimes $N_{ij}$ is even used in place of the inflaton field to describe observables such as the spectral index $n_s$ and the tensor-to-scalar ratio $r$. However, it is important to notice that to date there is no clear and direct way of obtaining bounds for $N_{ij}$ that are independent of a model of inflation.

\begin{figure}[tb]
\captionsetup{justification=raggedright,singlelinecheck=false}
\par
\includegraphics[trim = 0mm  0mm 1mm 1mm, clip, width=8.9cm, height=5.43cm]{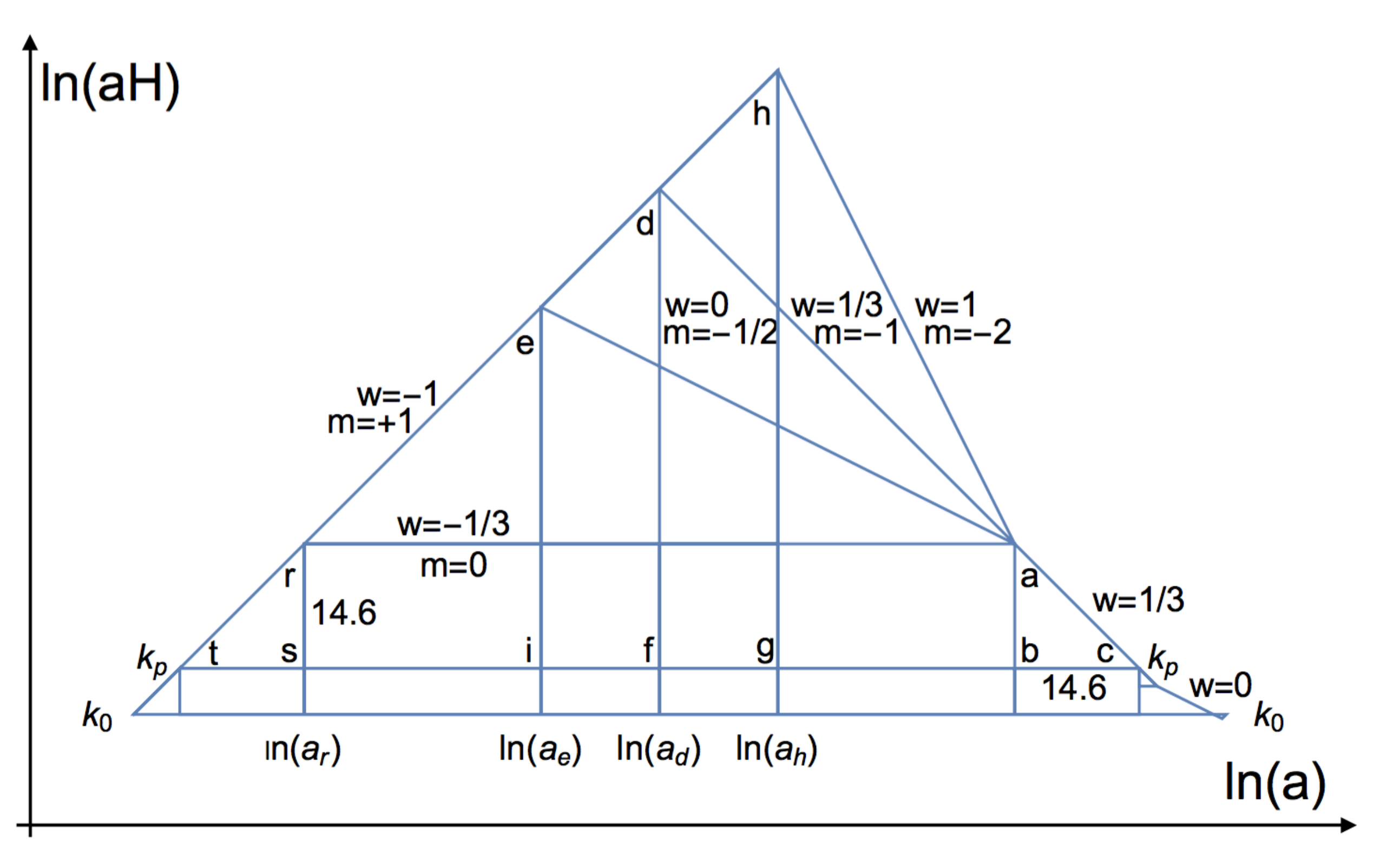}
\caption{\small Evolution of the logarithm of the comoving Hubble scale wavenumber mode  $\ln(k=a\,H)$ as a function of $\ln(a)$. The comoving scale $k=aH$ exits the horizon during inflation (line in the l.h.s of the figure of slope $m=+1$) and reenters  at the pivot scale $k_p=a_pH_p$ (line in the r.h.s of the figure of slope $m=-1$) during radiation domination. We know that the scale reenters at $k_p$ during the radiation era because at the scale $k_p$ the scale factor is $a_p < a_{eq}$, where $a_{eq}$ is the scale factor at matter-radiation equality. Lines for the $\omega=-1/3$, $\omega=0$, $\omega=1/3$ and $\omega=1$ cases are also drawn. By measuring their projection on the $\ln(a)$ axis with a ruler we can determine bounds for the number of e-folds as shown in Table~\ref{Nbounds}. The reader can easily draw her/his own figure by remembering that the fixed main framework is given by the two big triangles $\overline {tfd\,t}$ and $\overline {cfd\,c}$ which have the same dimensions thus, $\overline {tf}$=$\overline {fd}$=$\overline {fc}$.
}
\label{Diagrama01}
\end{figure}

In this article we present a simple geometrical approach which allows the determination of $N_{ij}$ directly by measuring lengths of segments in Fig.~\ref{Diagrama01}. The strategy is based in two simple observations: that it is possible to obtain an upper bound to the total number of e-folds of expansion of the universe from $a_k$ during inflation to the pivot scale with wavenumber mode $k=k_p=\frac{0.05}{Mpc}$ or to the present scale $k=k_0$ (see Eq.~\eqref{EQbound} below) and that the Hubble line specifying the radiation era with $\omega=1/3$ and slope $m=-1$ is fixed by determining the pivot scale factor $a_p$ at $k=k_p$. The calibration for further measurements is simply obtained by dividing the total number of e-folds of expansion given by Eq.~\eqref{EQbound}  over the length of the corresponding segment; this gives the number of e-folds per unit length.

The outline of the paper is as follows: in Section \ref{DIA} we discuss the diagram shown in Fig.~\ref{Diagrama01} and explain how can we possibly measure $N_{ij}$ directly from the figure. At the end of the section we give a lower bound to the size of the  universe at the beginning of observable inflation. Section \ref{BOU} contains our measurements for the  $-1/3<\omega<1$ and $0<\omega<1/3$ cases which are displayed in two pairs of columns in the Table~\ref{Nbounds}. Section~\ref{POS} presents a $postdiction$ where consistency with a relevant diagram requires the existence of a new form of energy to drive the present expansion of the universe. Section~\ref{CON} contains the main conclusions on the results presented in the article as well as a further discussion of possible extensions and applications of our approach. Finally, a collection of definitions for the various $N_{ij}\equiv \ln\frac{a_j}{a_i}$ used throughout the article is given in the Appendix.
\section {\bf The diagram}\label{DIA}
Let us briefly explain the general strategy for obtaining bounds to the number of e-folds. To do this it is necessary to draw a fixed framework containing and constraining the epochs of reheating and of radiation (see Fig.~\ref{Diagrama01}). The line representing inflation is given by a straight line of slope $m=+1$ in the diagram showing $\ln(k)$ versus $\ln(a)$ where $k\equiv aH$ and $a$ is the scale factor, the inflationary line cuts the $\ln(a)$ axis at $\ln(a_k)$. On the r.h.s of the figure we have a line with slope $m=-1$ which represents radiation according to the equation relating the slope with the equation of state parameter (EoS)
\begin{equation}
m=-\frac{1+3\omega}{2}\;.
\label{m}
\end{equation}
This second line cuts the $\ln(a)$ axis at $\ln(a_p)$ where $a_p$ is the scale factor at the pivot scale wavenumber $k_p$. To calculate $a_p$ we proceed as follows: we calculate the scales in Planck mass units  where $M_{pl}=2.4357\times 10^{18} GeV$  and then set $M_{pl}=1$. Taking $h=0.674$ we get $k_0= 5.8925 \times 10^{-61}$ also $k_p=\frac{0.05}{Mpc}=1.3105\times 10^{-58}$. 
The value of the vertical ``distance" between the scales $k_p$ and $k_0$ (see Fig.~\ref{Diagrama01}) is
\begin{equation}
\ln\left(\frac{k_p}{k_0}\right)=5.4\;,
 \label{Frid}
\end{equation}
and is clearly the same in both sides of the figure. This does not occur, however, with the horizontal ``distances". While in the l.h.s triangle it is also equal to 5.4 e-folds the corresponding r.h.s distance is $\ln(a_0/a_p)$, larger than 5.4. To calculate $a_p$ we write the Friedmann equation in the form
\begin{equation}
H_p=H_0\sqrt{\frac{\Omega_{md,0}}{a_p^3}+\frac{\Omega_{rd,0}}{a_p^4}+\Omega_{de}}\;, 
 \label{Frid}
\end{equation}
from where it follows that
\begin{equation}
\ln\left(\frac{k_p}{k_0}\right)=\frac{1}{2}\ln\left(\frac{\Omega_{md,0}}{a_p}+\frac{\Omega_{rd,0}}{a_p^2}+\Omega_{de}a_p^2\right)\;, 
 \label{EQas}
\end{equation}
where $\Omega_{md,0}=0.315,$ $\Omega_{rd,0}=5.443\times 10^{-5},$ and $\Omega_{de}=0.685$. The resulting solution to Eq.~(\ref{EQas}) is $a_p=3.6512\times10^{-5}$ and the number of e-folds from $a_p$ to $a_0$ is $N_{p0}=\ln(a_0/a_p)=10.2$.
Since we find that $a_p < a_{eq}$ where $a_{eq}$ is the scale factor at matter-radiation equality, the radiation line is fixed. To fix the inflation line we have to fix $a_k$, to do this we use an equation recently proposed by the author \cite{German:2020dih} 
\begin{equation}
N_{ke}+N_{ep}=\ln[\frac{a_{p}H_k}{k_p}]\;,
\label{EQ}
\end{equation}
where $N_{ke}\equiv\ln(\frac{a_e}{a_k})$ is the number of e-folds from $a_k$ to the end of inflation at $a_e$ and $N_{ep}\equiv\ln(\frac{a_p}{a_e})$ is the number of e-folds from the end of inflation to  the pivot scale at $a_p$. The Hubble function can be written as $H=\pi\sqrt{ r A_s/2}$ and using the bound $r<0.063$, \cite{Aghanim:2018eyx}, \cite{Akrami:2018odb}  at $k_p$  it follows that
\begin{equation}
N_{ke}+N_{ep}=\ln[\frac{a_{p}\pi\sqrt{ A_s}}{\sqrt{2}\,k_p}]+\frac{1}{2}\ln\,r<112.5\;,
\label{EQbound}
\end{equation} 
where $A_s=2.1\times 10^{-9}$ is the scalar power spectrum amplitude.  Fixing $N_{ke}+N_{ep}$ at the bound fixes $a_k$ and the inflationary line. Thus, we have built a framework on which to refer all other lines of interest. By dividing 112.5 over the length of the segment $\overline {tc}$ we obtain a bound to the number of e-folds per unit length. From there we can determine bounds for the number of e-folds for any equation of state $\omega$  by direct measurement on Fig.~\ref{Diagrama01}. In the diagrammatic approach presented here the Hubble function $H$ during inflation is actually constant (de Sitter universe) while in models of inflation $H_k$ is not. However, the departures from constant $H$ are minute as shown in  \cite{German:2020dih} for the Starobinsky model of inflation where maximum departures amount to $0.1\%$; a mere 0.1 e-fold .

\begin{figure}[tb]
\captionsetup{justification=raggedright,singlelinecheck=false}
\par
\includegraphics[trim = 0mm  0mm 1mm 1mm, clip, width=8.85cm, height=5.3cm]{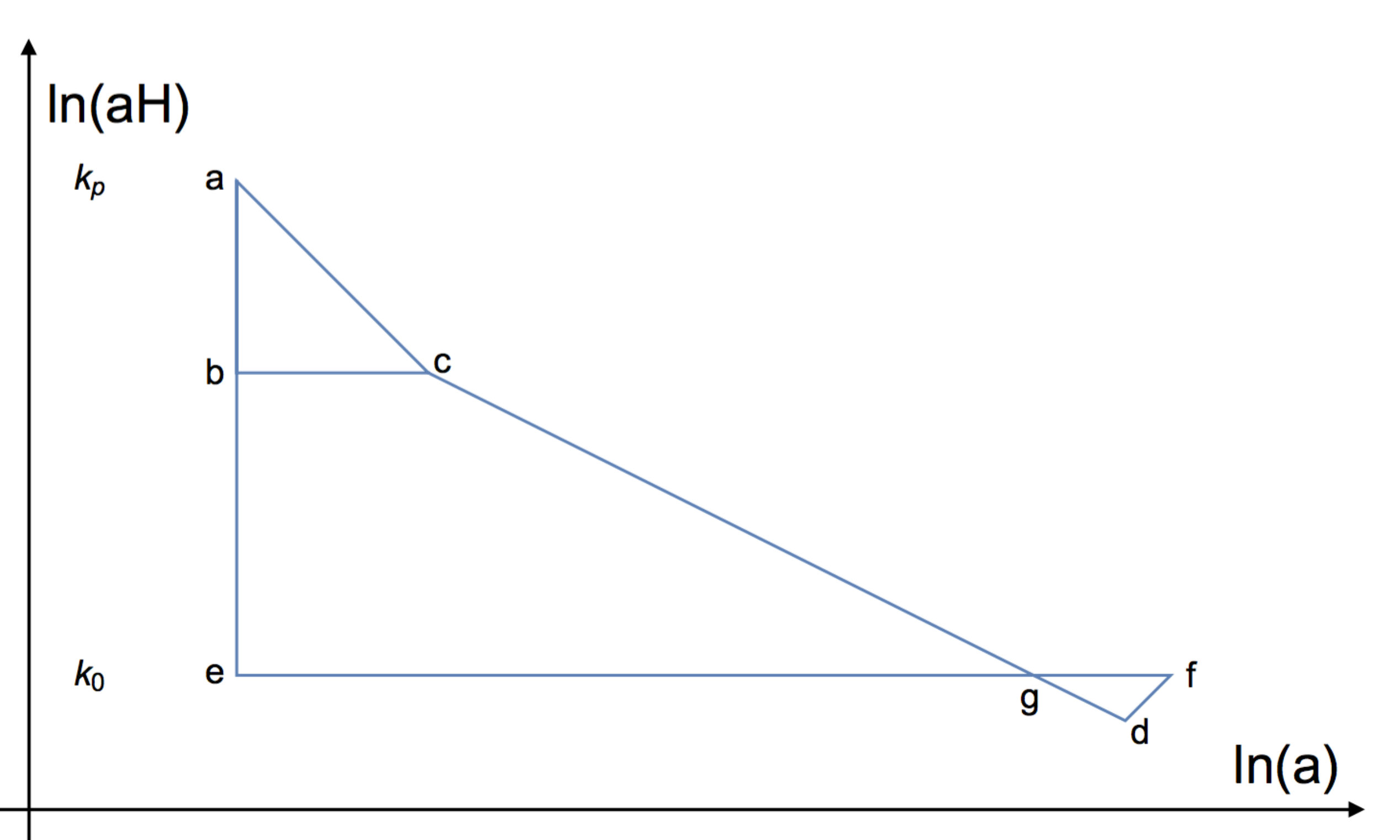}
\caption{\small The figure shows an amplification of the r.h.s. corner of Fig.~\ref{Diagrama01}. The segment $\overline {ac}$ corresponds to the last 2.1 e-folds of the radiation dominated era. The segment $\overline {cd}$ describes the 7.6 e-folds of matter domination and the last bit $\overline {df}$ the 0.5 e-folds of dark energy domination up to the present at $f$. The scale factors at $a$, $c$, $d$ and $f$ are $a_p=3.65\times10^{-5}$, $a_{eq}=2.97\times10^{-4}$, $a_{de}=0.61$ and $a_{0}=1$, respectively.
}
\label{Diagrama01c}
\end{figure}

To calculate the bounds it is necessary to know the $minimum$ number of e-folds during radiation. The number of e-folds during the radiation dominated era is given by $N_{rd}=\ln\left(\frac{T_{r}}{(\frac{4}{11})^{1/3}T_0}\right)-\ln(\frac{a_0}{a_{eq}})$ where $T_{r}$ is the temperature of the universe at the beginning of radiation, $T_0=2.725K$ and $a_{eq}=2.97\times 10^{-4}$. Also, the number of e-folds during the radiation dominated era from the beginning of radiation at  $a_{r}$ to the pivot scale factor $a_p$ is given by $N_{rp}\equiv \ln(\frac{a_{p}}{a_{r}})=\ln\left(\frac{T_{r}}{(\frac{4}{11})^{1/3}T_0}\right)-\ln(\frac{a_0}{a_{p}})$. We see that both numbers depend on the temperature $T_{r}$ however their difference $N_{peq}\equiv \ln(\frac{a_{eq}}{a_p})=N_{rd}-N_{rp}$ does not. 
We find that $N_{peq}=2.1$ e-folds (the very small triangle just below $k_p$ in the r.h.s of the diagram, amplified in Fig.~\ref{Diagrama01c}). The important point is that $a_p < a_{eq}$ and $k_p$ is inside the radiation dominated era. Thus, the radiation line is represented in the diagram of Fig.~\ref{Diagrama01} by the segment $\overline {ac}$ with $\omega=1/3$ and slope $m=-1$ as well as by the extra bit (also denoted $\overline {ac}$) in Fig.~\ref{Diagrama01c}. 
The minimum possible temperature $T_{r}$ is bounded by nucleosynthesis requirements. For definitiveness we choose $T_{r}=10 MeV$  and this gives, for the segment $\overline {bc}$, $N_{rp}=14.6$ e-folds. There are still $2.1$ e-folds from $a_p$ to the end of radiation at $a_{eq}$ (point $c$ in Fig.~\ref{Diagrama01c}), $7.6$ e-folds of matter domination (segment $\overline {c\,d}$ in Fig.~\ref{Diagrama01c}) and a further 0.5 e-folds of dark energy domination (segment $\overline {df}$ in Fig.~\ref{Diagrama01c}). Thus, there are $N_{p0}\equiv \ln(\frac{a_0}{a_p})=10.2$ e-folds from $a_p$ to the present. If we had chosen a different value for $T_{r}$, let us say $T_{r}=1\,MeV$ then $N_{rp}$ would not be 14.6 but 12.3 this would change the size of the triangle $\overline {abc}$ and thus, the bounds for the number of e-folds during reheating and during radiation quoted below but not significantly. However $N_{peq}$ and $N_{p0}$ would remain the same. 

 Before proceeding with our measurements, just a small comment in relation with the size of the universe at the beginning of observable inflation. From the discussion above we see that there is an upper bound for the total number of e-folds for the observable universe given by measuring the bottom $k_0$ line or by adding the extra e-folds to the $k_p$ line: 112.5+5.4+10.2=128.1 e-folds where 5.4 comes from the expansion between $\ln(k_0)$ to $\ln(k_p)$  during inflation (l.h.s. triangle with lower vertex at $\ln(k_0)$ in Fig.~\ref{Diagrama01}) and 10.2 comes from the number of e-folds from $a_p$ to $a_0$ (r.h.s. composition of triangles with lower vertex at $\ln(k_0)$ in Fig.~\ref{Diagrama01}). From here we find that the scale factor at the beginning of the inflation giving rise to our observable universe is $a_{k_0}=e^{-128.1}a_0=2.3\times 10^{-56}$. For our present universe of diameter $8.8 \times 10^{26}m$ we find that the size of the early universe at the beginning of observable inflation at $a_{k_0}$ was at least $2.05 \times 10^{-29}m$ in diameter, that is, at least $1.27 \times 10^{6}$ times larger than the Planck length.
Finally we would like to remark that all results presented in this article are model independent, in the sense that no model of inflation has been used to obtain them.
\begin{table*}[htbp!]
\captionsetup{justification=raggedright,singlelinecheck=false}
\par
\caption{\label{Nbounds}  In the Appendix we give the definitions for the various $N_{ij}$ shown in the table. We display in the first row bounds for the total number of e-folds from inflation $plus$ post inflation evolution for the pivot scale wavenumber $k=k_p=\frac{0.05}{Mpc}$ as well as for the present scale at $k=k_0=a_0H_0$. These bounds come from  Eq.~\eqref{EQbound} where a bound $r<0.063$ to the tensor-to-scalar ratio has been imposed.
Columns first and second display bounds for the cases $-1/3<\omega<1$, both for $k=k_p$ and $k=k_0$. Columns third and fourth show bounds for $0<\omega<1/3$, for the $k=k_p$ and $k=k_0$ cases. The number of e-folds during the radiation era is $N_{req}\equiv\ln\left(\frac{a_{eq}}{a_r}\right)$ is bounded as $16.7<N_{req}<58.3$ . In our definition of $N_{req}$ just given, $a_{r}$ is the scale factor at the beginning of the radiation era and $a_{eq}$ is the matter-radiation equality scale factor, }
\begin{ruledtabular}
\begin{tabular}{cccc}
$-1/3<\omega<1,\quad k=k_p$ & $-1/3<\omega<1,\quad k=k_0$ & $0<\omega<1/3,\quad k=k_p$ & $0<\omega<1/3,\quad k=k_0$ \\[2mm]
 \hline\\[0.2mm]
$N_{ke}+N_{ep}<112.5$ & $N_{ke}+N_{e0}<128.1$ & $N_{ke}+N_{ep}<112.5$ & $N_{ke}+N_{e0}<128.1$\\[2mm]
$14.6<N_{ke}<70.1$ & $20<N_{ke}<75.5$ & $42.4<N_{ke}<56.2$ & $47.8<N_{ke}<61.6$\\[2mm]
$97.9>N_{ep}>42.4$  & $108.1>N_{e0}>52.6$ & $70.1>N_{ep}>56.2$ & $80.3>N_{e0}>66.5$\\[2mm]
 $83.3>N_{er} \geq 0$   & $83.3>N_{er} \geq 0$ & $55.5>N_{er} \geq0$ & $ 55.5>N_{er} \geq0$\\[2mm]
$14.6<N_{rp}<56.2$   & $24.8<N_{r0}<66.4$ & $14.6<N_{rp}<56.2$ & $24.8<N_{r0}<66.4$ \\[2mm]
 \end{tabular}
\end{ruledtabular}
\end{table*}
\section {\bf Bounds for the number of e-folds}\label{BOU}
\subsection {\bf The case $-1/3<\omega<1$}\label{BOUA}
First we will determine bounds when $\omega=-1/3$ (line of slope $m=0$ in Fig.~\ref{Diagrama01}). This line (segment $\overline {ra}$), when composed with the smallest number of e-folds from radiation (segment $\overline {bc}$), gives the most $post$ $inflationary$ e-folds. This means a lower bound for  the number of e-folds during inflation $N_{ke}$, from $a_k$ at $k=k_p$ to the end of inflation at $a_{r}$ (segment $\overline {ts}$). The task is then very simple: we take our ruler and measure e.g., in millimeters the length of the segment $\overline {ts}$, multiply it by the calibration obtained before (after Eq.~\eqref{EQbound}) and the resulting number is the minimum number of e-folds during inflation: $14.6<N_{ke}$ (see first column of Table~\ref{Nbounds}). To obtain an upper bound for $N_{ke}$ we look at the line with $\omega=1$ and slope $m=-2$ (segment $\overline {ah}$) which when composed with the smallest number of e-folds from radiation (segment $\overline {bc}$), gives the least post inflationary e-folds. This means an upper bound for  the number of e-folds during inflation $N_{ke}$, from $a_k$ at $k=k_p$ to the end of inflation at $a_{h}$ (segment $\overline {tg}$). The resulting bounds for $N_{ke}$ at $k=k_p$ are $14.6<N_{ke}<70.1$ as shown in the first column of Table~\ref{Nbounds}. In the table also appear bounds for the number of post inflationary e-folds $N_{ep}$ and bounds for the reheating era $N_{er}$ all of them measured at the scale $k_p$. To obtain the corresponding bounds for the present scale $k_0$ we simply measure along the largest horizontal line at $k_0$ with the results presented in the second column of Table~\ref{Nbounds}. In particular we find that the number of e-folds during inflation is bounded as $20<N_{ke}<75.5$ at $k_0$, in excellent agreement with old estimates \cite{Liddle:2003as}, \cite{Dodelson:2003vq} ranging from 18 to 77 e-folds. Since this is really an article on experimental physics, uncertainties should be given. Thus we convert millimeters into e-folds with the result that all the numbers quoted in Table~\ref{Nbounds} should carry a $\pm\, 0.2$ e-folds but it is not written there for clarity of the presentation.

\subsection {\bf The case $0<\omega<1/3$}\label{BOUB}
Proceeding in a similar way to the previous case we now determine bounds when $0<\omega<1/3$. First we pay attention to the line with $\omega=0$ and slope $m=-1/2$ (segment $\overline {ae}$) in Fig.~\ref{Diagrama01}. The minimum number of e-folds during inflation at $k=k_p$ is given by the length of the segment $\overline {ti}$, in this case $42.4<N_{ke}$ with $70.1>N_{ep}$ and $55.5>N_{er}$. The upper bound for $N_{ke}$ occurs for the very symmetrical case $\omega=1/3$ where we have that $N_{ke}=N_{ep}=56.2$. Results for $0<\omega<1/3$ at $k=k_p$ are given in the third column of Table~\ref{Nbounds} with those corresponding to the present scale wavenumber $k_0$ in the fourth column. 

Note that for a given $\omega<1/3$ the maximum $N_{ke}$ occurs when we displace the segment e.g., $\overline {ea}$ in the  $\omega=0$ case parallel to itself along the radiation line i.e., we slide the vertex at $a$ towards $d$ with $b$ of the segment $\overline {c\, b}$ approaching $f$. In the limit $a$ reaching $d$ we get 0 e-folds for reheating and a maximum number of e-folds for the radiation era (instantaneous reheating and maximum temperature). Thus the maximum number of e-folds for $N_{ke}$ from $a_k$ to $a_{d}$ (segment $\overline {tf}$) is 56.2 at $k=k_p$ for any $\omega<1/3$ (see  \cite{German:2020dih}  for a concrete example using Starobinsky model of inflation). 
\section {\bf The postdiction}\label{POS}
\begin{figure}[tb]
\captionsetup{justification=raggedright,singlelinecheck=false}
\par
\includegraphics[trim = 0mm  0mm 1mm 1mm, clip, width=8.85cm, height=5.20cm]{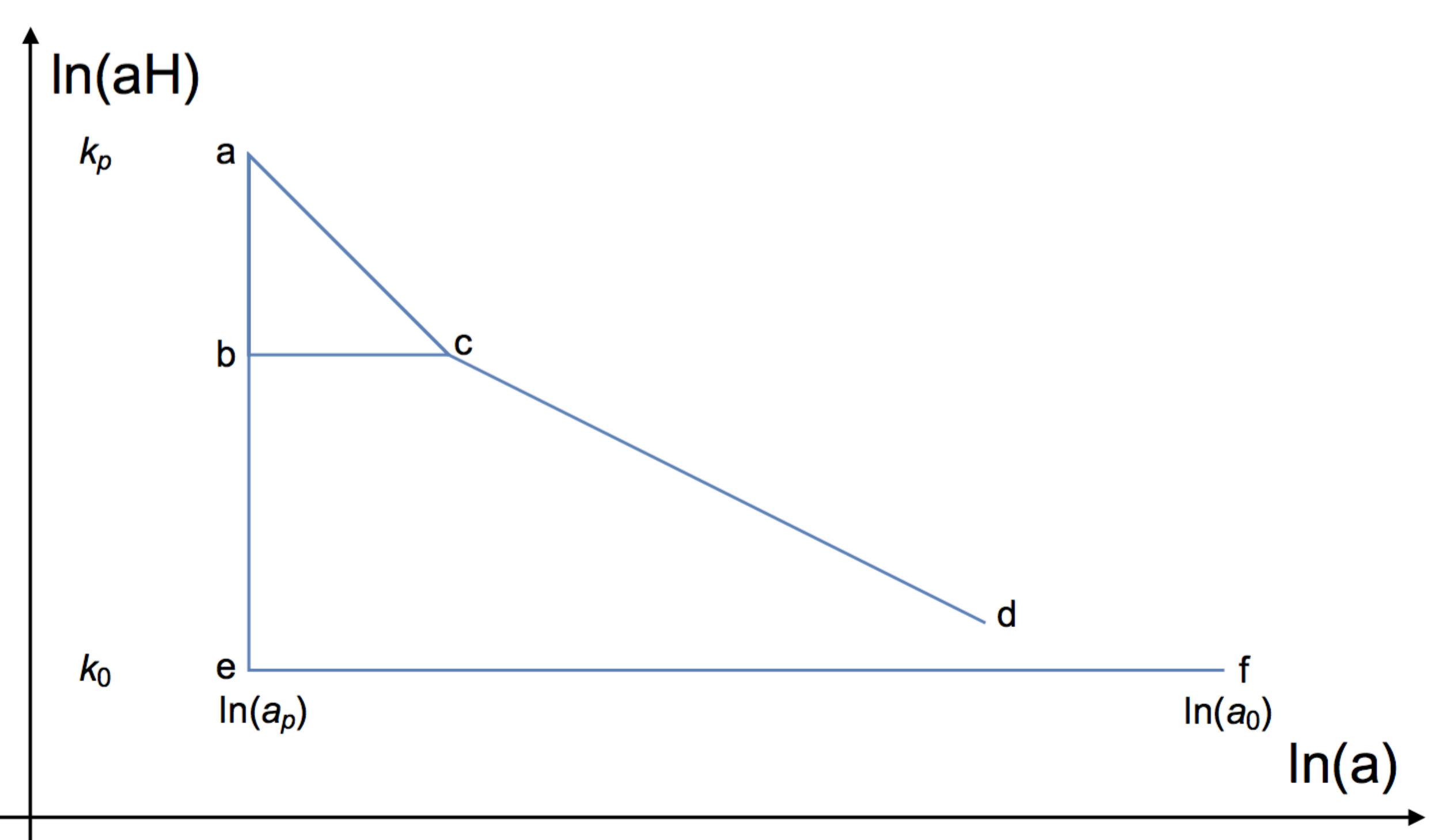}
\caption{\small Assuming that we have not heard about the dark energy the r.h.s. corner of Fig.~\ref{Diagrama01} would have looked like this. In our effort to connect the (arbitrarily chosen) point at $d$ with the present at $f$ we would immediately see that we could not reach $f$ by continuing the segment $\overline {cd}$ with the same slope. We would have to draw a new line with a different slope (different EoS) describing a new form of energy.
}
\label{Diagrama01a}
\end{figure}
While the previous description is useful for obtaining bounds on the number of e-folds of inflation, reheating  and radiation, it also allows us to study the regimen between the pivot scale and the present as illustrated in Fig.~\ref{Diagrama01c}. This diagram corresponds to the lower r.h.s corner of the main diagram (Fig.~\ref{Diagrama01}) and comprises the last bit of radiation $\overline {ac}$, the epoch of matter domination $\overline {c\,d}$ and the present dark energy dominated era $\overline {df}$ (see Fig.~\ref{Diagrama01c}).

\begin{figure}[tb]
\captionsetup{justification=raggedright,singlelinecheck=false}
\par
\includegraphics[trim = 0mm  0mm 1mm 1mm, clip, width=8.9cm, height=5.17cm]{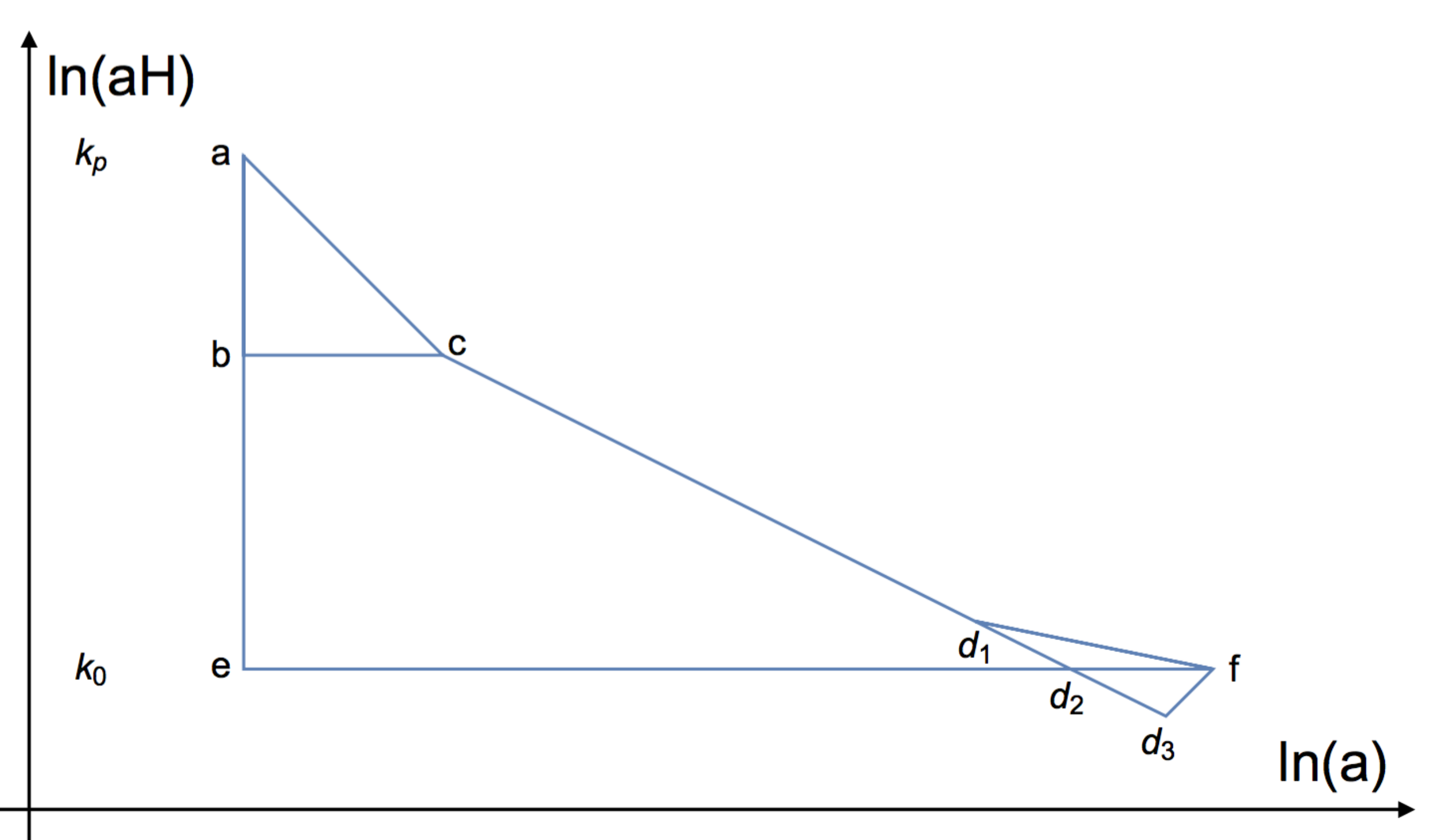}
\caption{\small The diagram shows three possible ways of joining the point $d$ to $f$ each with a given slope $m$ and corresponding EoS $\omega$. Lines  below the segment $\overline {d_2\,f}$  would have slope $m > 0$ and corresponding EoS $\omega < -1/3$ while those above would have slope $0>m>-1/2$ and corresponding EoS $-1/3<\omega<0$. One can show that $\omega < -0.062$ or that $m>-0.41$.  }
\label{Diagrama01b}
\end{figure}

In this section we would like to show the usefulness of the diagrammatic approach by elaborating on a $postdiction$ related to the necessity of a new form of energy between matter domination and the present. We will show that consistency with the relevant diagram requires the existence of a new form of energy to drive the present expansion of the universe.
Let us assume for a moment that we have no knowledge of the existence of dark energy as we know it today. We can assume that it is the year 1997 or some time before the first supernovae observations indicating a new form of energy driving the expantion of the universe where reported \cite{Perlmutter:1998np}, \cite{Riess:1998cb}. Thus we focus in the r.h.s lower section of Fig.~\ref{Diagrama01} ending in the present scale $k_0$. This section of the figure is amplified in Fig.~\ref{Diagrama01c}. But because we are assuming no observational evidence of the dark energy we  would rather have drawn the (incomplete) diagram shown in Fig.~\ref{Diagrama01a}, that is, we would be able to draw the segment $\overline {c\,d}$ with slope $-1/2$ and EoS $\omega=0$ corresponding to matter domination and we would be able to draw the segment $\overline {ef}$ corresponding to the 10.2 e-folds of expansion from $a_p$ to the present $a_0$. That we would know for sure, but in the process of joining the line ending in (some arbitrarily chosen point) $d$ to the present at $f$ we would realize that it is not possible to do it just by continuing the segment $\overline {c\,d}$ with the same slope (same EoS). 
We would have to change the slope of the line i.e., we would have to use an EoS different from $\omega=0$ and {\it this means that at some time during matter domination a new form of energy should overtake matter and drive the expansion of the universe}. This is shown in the diagram of Fig.~\ref{Diagrama01b} where three possible examples are given. Lines  below the segment $\overline {d_2\,f}$ (for example the segment $\overline {d_3\,f}$) would have slope $m > 0$ and corresponding EoS $\omega < -1/3$ while those above the segment $\overline {d_2\,f}$ (for example the segment $\overline {d_1\,f}$) would have slope $0>m>-1/2$ and corresponding EoS $-1/3<\omega<0$. Thus the new form of energy would have an EoS bounded as $\omega<0$ and slope $m> -1/2$; it could not be matter nor radiation. The diagram does not tell us what form of energy there should be but it does tell us (in a very simple way) that it can not be matter and that $\omega$ should be less than 0 (actually $\omega<-0.062$ in the very extreme situation that the new form of energy domination started at $c$). 

Before the point at $d$ we had matter domination, radiation domination and the inflation most likely by the dominance of the potential energy of a scalar field. Unless an exotic substance emerges during the evolution of the universe and because the last period of expansion to the present can not be matter or radiation dominated because their corresponding lines cannot join the matter line with the point at $f$ we would bet for the scalar field and late inflation but this wisdom is suspicious in the year 2020.
As it turned out that form of energy (dark energy) has an EoS $\omega\approx -1$ and slope $m\approx1$ and it is shown in the diagram of Fig.~\ref{Diagrama01c} by the segment $\overline {df}$. From there we can $measure$ that the number of e-folds of dark energy domination is $0.5$ from where we get a scale factor $a_{de}=0.61$ and redshift $z_{de}=0.64$. We can also see from Fig.~\ref{Diagrama01c} that scales the size of the present horizon had entered at $g$ (scale factor $a_g=0.22$, redshift $z_g=3.5$) but they have already left today (at $f$). 
\section {\bf Discussion and conclusions}\label{CON}
We have suggested a simple geometrical approach to the problem of determining bounds for the number of e-folds $N_{ij}$ for the various epochs of evolution of the universe. The main ingredients are the (model independent) bound given by Eq.~\eqref{EQbound} and the realisation that the radiation line ($\omega=1/3$, slope $m=-1$) is fixed by the pivot scale factor $a_p$ lying in the radiation era, (see Fig.~\ref{Diagrama01}). The scale to determine $N_{ij}$ is simply obtained by dividing the upper bound $N_{ke}+N_{ep}=112.5$ given by Eq.~\eqref{EQbound} over the length of the corresponding segment $\overline {tc}$ at $k=k_p$ . This gives the number of e-folds per unit length. With this calibration we determine $N_{ij}$ by measuring with a ruler the length of the segment in question, (see Table~\ref{Nbounds}). Work in progress indicate that this strategy is particularly useful when studying specific models of inflation and provides a simple picture from where to extract qualitative as well as quantitative information in an almost effortless way  \cite{German:2020dih} . 
We  mention that this approach can be improved in several ways. It is clear that the de Sitter assumption (that inflation is perfectly exponential or, equivalently, that the Hubble function is constant during inflation) is in general not satisfied by models of inflation although we expect that realistic corrections do not significantly alter our results  \cite{German:2020dih} . More importantly is the imposition of extra constraints (for a recent example see \cite{Tanin:2020qjw}) not yet present in our approach. In any case we believe the strategy presented here is useful and very simple, can be extended in several interesting directions (at the end of the Section \ref{DIA} we give a lower bound to the diameter size of the universe at the beginning of observable inflation) and can be applied to any model of inflation.

We have also shown by means of a simple figure that the need of a new form of energy follows from the requirement to connect the diagram of Fig.~\ref{Diagrama01a}. The line representing matter domination should join the present scale and it is clear that this can not be achieved by simply continuing the matter line to the present.  At least a new line with a different equation of state parameter $\omega$ should be drawn to overtake matter and drive the present expansion of the universe. Although not much value should be attached to this $postdiction$ it is clear from the previous discussion that the strategy used is a good indication of the usefulness of the diagrammatic approach. Finally, let us notice that what we have done in the last part of this letter is what we usually attempt when studying reheating: trying to join a line from an inflationary to a radiation line. There the problem has an extra degree of ignorance because we do not know where inflation ends (in the $\ln(a)$, $\ln(k)$ diagram) nor where radiation starts.
\section*{Acknowledgements}
It is a pleasure to thank N. S\'anchez and M. Dirzo for encouraging conversations. We acknowledge financial support from UNAM-PAPIIT,  IN104119, {\it Estudios en gravitaci\'on y cosmolog\'ia}.
\\
\\
\appendix{\bf Appendix: Defining the number of e-folds $N_{ij}$ }\label{APP}
\\
\\
Below and for easy reference we define the notation for the various number of e-folds $N_{ij}$ discussed in the article, where $N_{ij}\equiv \ln\frac{a_j}{a_i}$ measures the expansion of the universe from the time the scale factor was $a_i$ to the time when it was $a_j$. \\[2mm]
$N_{ke}+N_{ep}=\ln\left(\frac{a_p}{a_k}\right)$,   from $a_k$ during inflation to the pivot scale factor  $a_p$.\\[0.1mm]
$N_{ke}+N_{e0}=\ln\left(\frac{a_0}{a_k}\right)$,   from $a_k$ during inflation to the present scale factor $a_0$.\\[0.1mm]
$N_{ke}\equiv\ln\left(\frac{a_e}{a_k}\right)$,   from $a_k$ during inflation to the end of inflation at $a_e$.\\[0.1mm]
$N_{ep}\equiv\ln\left(\frac{a_p}{a_e}\right)$, from the end of inflation to the pivot scale factor $a_p$.\\[0.1mm]
$N_{e0}\equiv\ln\left(\frac{a_0}{a_e}\right)$, from the end of inflation to the present at $a_0$.  \\[0.1mm]
$N_{er}\equiv\ln\left(\frac{a_{r}}{a_e}\right)$,   from the end of inflation to the beginning of radiation.\\[0.1mm]
$N_{rp}\equiv\ln\left(\frac{a_{p}}{a_r}\right)$,   from the beginning of radiation to the pivot scale factor $a_p$.\\[0.1mm]
$N_{r0}\equiv\ln\left(\frac{a_{0}}{a_r}\right)$,   from the beginning of radiation to the present scale factor $a_0$.\\[0.1mm]
$N_{req}\equiv\ln\left(\frac{a_{eq}}{a_r}\right)$,   from the beginning of radiation to the matter-radiation equality.\\[0.1mm]
\\
\\
\\

\end{document}